# Higher Order Binaries with Time Dependent Coefficients and Two Factors - Model for Defaultable Bond with Discrete Default Information

**Hyong-Chol O** , **Yong-Gon Kim** and **Dong-Hyok Kim**

*Faculty of Mathematics, **Kim Il Sung** University, Pyongyang, D. P. R. Korea*

*Abstract*: In this article, we consider a 2 factors-model for pricing defaultable bond with discrete default intensity and barrier where the 2 factors are stochastic risk free short rate process and firm value process. We assume that the default event occurs in an expected manner when the firm value reaches a given default barrier at predetermined discrete announcing dates or in an unexpected manner at the first jump time of a Poisson process with given default intensity given by a step function of time variable. Then our pricing model is given by a solving problem of several linear PDEs with *variable coefficients* and terminal value of binary type in every subinterval between the two adjacent announcing dates. Our main approach is to use higher order binaries. We first provide the pricing formulae of *higher order binaries* with *time dependent coefficients* and consider their integrals on the last expiry date variable. Then using the pricing formulae of higher binary options and their integrals, we give the pricing formulae of defaultable bonds in both cases of exogenous and endogenous default recoveries and credit spread analysis.

## 1. Introduction

The study on defaultable corporate bond and credit risk is now one of the most promising areas of cutting edge in financial mathematics [1]. As well known, there are two main approaches to pricing defaultable corporate bonds; one is the *structural approach* and the other one is the *reduced form approach*. In the structural method, we think that the default event occurs when the firm value is not enough to repay debt, that is, the firm value reaches a certain lower threshold (*default barrier*) from the above. Such a default can be expected and thus we call it *expected default*. In the reduced-form approach, the default is treated as an unpredictable event governed by a default intensity process. In this case, the default event can occur without any correlation with the firm value and such a default is called *unexpected default*. In the reduced-form approach, if the default probability in time interval $[t, \ t + \Delta t]$ is $\lambda \Delta t$, then $\lambda$ is called *default intensity* or *hazard rate*.

As for the history of the two approaches and their advantages and shortcomings, readers can refer to the introductions of [3, 6, 7, 9, 15]. To take the advantages and overcome the shortcomings of structural and reduced-form approaches, many authors used unified models of the two approaches. (See [3, 4, 6, 7, 9, 10, 13, 14, 15, 16].) As noted in [10, 13, 14], many researchers of unified model including [3, 4, 6, 7, 9, 15] tried to express the price of the bond in terms of the firm value or the related signal variable to the firm value and the value of default intensity together with default barrier at any time in the whole lifetime of the bond.

On the other hand, in [13, 10] the author noted that it is difficult for investors outside of the firm to know the firm's financial data except for some discrete dates (for example, once in a month or once in a three month etc.) to announce management data and studied the pricing problem for defaultable corporate bond under the assumption that we only know the firm value and the default barrier at 2 fixed discrete announcing dates, we don't know about any information of the firm value in another time and the default intensity between the adjoined two announcing dates is a constant determined by its announced firm value at the former announcing date. The computational error in [13] is corrected in [10]. The approach of [13, 10] is a kind of study of defaultable bond under insufficient information about the firm. It is interesting to note that Agliardi et al [2] studied bond pricing problem under *imprecise information* with the technique of fuzzy mathematics. The approach of [13, 10] can also be seen as a *unified model* of structural model and reduced form model. Agliardi [1] studied a *structural model* for defaultable bond with several (discrete) coupon dates where the default can occur only when the firm value is not large enough to pay its debt and coupon in those discrete coupon dates.

In [14], the authors studied *one-factor model* for defaultable bond with discrete default intensity and discrete default barrier using higher order binary options and their integrals, where the 1 factor is the firm value process. In their credit risk model, the default event occurs in an expected manner when the firm value reaches a certain lower threshold - the default barrier at predetermined discrete announcing dates or in an unexpected manner at the first jump time of a Poisson process with given default intensity given by a step function of time variable, respectively. They considered both endogenous and exogenous default recovery and the pricing model is a solving problem of inhomogeneous or homogeneous Black-Scholes PDEs with different coefficients and terminal value of binary type in every subinterval between the two adjacent announcing dates. In order to deal with the inhomogenous term related to endogenous recovery, they introduced *a special* binary option called *integral of i-th binary or nothing* and using it obtained the pricing formulae of defaultable corporate bond. The approach of [14] to model credit risk seems similar with the one of [10] but the essential difference is that in [10] they assumed that they know the firm value only in the discrete announcing dates and the default intensity between two adjacent announcing dates is determined by the firm value in the former announcing date. Another different point is that [14] considered arbitrary number of announcing dates but [10] considered only 2 announcing dates.

As a continued study of [14] we here consider a *two factors - model* for pricing defaultable bond with discrete default intensity and barrier where the 2 factors are stochastic risk free short rate process and firm value process. Our pricing model is given by a solving problem of several PDEs with *variable coefficients* and terminal value of binary type in every subinterval between the two adjacent announcing dates. Through the change of numeraire, they are transformed into several homogeneous or inhomogeneous Black-Scholes PDEs with *different time dependent coefficients* and terminal value of *binary type*. The coefficients time dependency is the *different point* from [14]. Here we encounter the problems of *higher order binaries* with *time dependent coefficients* even if the drifts and volatilities of short rate and firm value processes are all constants. Therefore we first provide the pricing formulae of higher order binaries with time dependent coefficients and consider their integrals on the last expiry date variable. Then using the pricing formulae of higher binary options and their integrals, we give the pricing formulae of defaultable bonds in both cases of exogenous and endogenous default recoveries and credit spread analysis.

Finally we note that it is interesting to see that the Geske's compound option approach used in [1] for pricing of defaultable bond with discrete coupon payments in structural approach is the same technique as higher binary used here.

The remainder of the article is organized as follows. In section 2 we consider higher order binaries with time dependent coefficients and their properties. In section 3 we set the problem for defaultable bonds and provide the pricing formulae and credit spread analysis. In section 4 we provide the sketch of the proof of pricing formulae for defaultable bonds.

## 2.    Higher Order Binaries with Time Dependent Coefficients

First, we explain higher order bond and asset binaries with risk free rate $r(t)$, dividend rate $q(t)$ and volatility $\sigma(t)$.

$$\frac{\partial V}{\partial t} + \frac{\sigma^2(t)}{2} x^2 \frac{\partial^2 V}{\partial x^2} + (r(t) - q(t))x \frac{\partial V}{\partial x} - r(t)V = 0, \quad 0 \le t < T, \ 0 < x < \infty, \qquad (2.1)$$

$$V(x,T) = x \cdot 1(sx > s\xi), \qquad (2.2)$$

$$V(x,T) = 1(sx > s\xi). \qquad (2.3)$$

The solution to the problem (2.1) and (2.2) is called the ***asset-or-nothing binaries*** (or **asset** binaries) and denoted by $A_{\bar{\xi}}^s(x,t\,;T)$ . The solution to the problem (2.1) and (2.3) is called the **cash-or-nothing binaries** (or **bond** binaries) and denoted by $B_{\bar{\xi}}^s(x,t\,;T)$ . Asset binary and

bond binary are called the **first order binary** options. If necessary, we will denote by

$A_{\bar{\xi}}^s(x,t\,;T\,;r(\cdot),q(\cdot),\sigma(\cdot))$  or $B_{\bar{\xi}}^s(x,t\,;T\,;r(\cdot),q(\cdot),\sigma(\cdot))$ , where the coefficients $r(t)$ , $q(t)$  and  $\sigma(t)$  of

Black-Scholes equation (2.1) are explicitly included in the notation.

   Let assume that $0 < T_0 < T_1 < \cdots < T_{n-1}$ and the $(n-1)$ th order (asset or bond) binary options  $A_{\xi_1\cdots\xi_{n-1}}^{s_1\cdots s_{n-1}}(x,t;T_1,\cdots,T_{n-1})$  and  $B_{\xi_1\cdots\xi_{n-1}}^{s_1\cdots s_{n-1}}(x,t;T_1,\cdots,T_{n-1})$ are already defined. Let

$$V(x,T_0) = A_{\xi_1\cdots\xi_{n-1}}^{s_1\cdots s_{n-1}}(x,T_0;T_1,\cdots,T_{n-1})\cdot 1(s_0 x > s_0\xi_0)\,, \qquad (2.4)$$

$$V(x,T_0) = B_{\xi_1\cdots\xi_{n-1}}^{s_1\cdots s_{n-1}}(x,T_0;T_1,\cdots,T_{n-1})\cdot 1(s_0 x > s_0\xi_0)\,. \qquad (2.5)$$

The solution to the problem (2.1) and (2.4) is called the **n-th order asset binaries** and denoted by $A_{\xi_0\xi_1\cdots\xi_{n-1}}^{s_0 s_1\cdots s_{n-1}}(x,t;T_0,T_1,\cdots,T_{n-1})$ . The solution to the problem (2.1) and (2.5) is called the

**n**-*th* **order bond binaries** and denoted by $B_{\xi_0\xi_1\cdots\xi_{n-1}}^{s_0 s_1\cdots s_{n-1}}(x,t;T_0,T_1,\cdots,T_{n-1})$ .

   Next, we provide the pricing formulae of asset and bond binaries with time dependent coefficients. To this end, we need the following proposition.

$$V(x,T) = f(x) \qquad (2.6)$$

   **Lemma 1**. *Assume that there exist nonnegative constants* $M$  *and*  $\alpha$  *such that* $|f(x)| \le$

$Mx^{\alpha\ln x}$, $x > 0$ . *Then the solution of* (2.1) *and* (2.6) *is provided as follows* :

$$V(x,T\,;T) = e^{-\bar{r}(t,T)}\int_0^\infty \frac{1}{\sqrt{2\pi\overline{\sigma^2}(t,T)}}\frac{1}{z}e^{-\frac{\left(\ln\frac{x}{z}+\bar{r}(t,T)-\bar{q}(t,T)-\frac{1}{2}\overline{\sigma^2}(t,T)\right)^2}{2\overline{\sigma^2}(t,T)}}f(z)dz$$

$$= xe^{-\bar{q}(t,T)}\int_0^\infty \frac{1}{\sqrt{2\pi\overline{\sigma^2}(t,T)}}\frac{1}{z^2}e^{-\frac{\left(\ln\frac{x}{z}+(\bar{r}(t,T)-\bar{q}(t,T)+\frac{1}{2}\overline{\sigma^2}(t,T)\right)^2}{2\overline{\sigma^2}(t,T)}}f(z)dz\,. \qquad (2.7)$$

*Here*

$$\bar{r}(t,T) = \int_t^T r(s)ds, \qquad \bar{q}(t,T) = \int_t^T q(s)ds, \qquad \overline{\sigma^2}(t,T) = \int_t^T \sigma^2(s)ds\,. \qquad (2.8)$$

   **Proof.** It is well known that the solution to Black-Scholes equation with time dependent coefficients $r(t), q(t)$  and  $\sigma(t)$ can be obtained by replacing $r(T-t), q(T-t)$  and  $\sigma^2(T-t)$ in the solution representation of Black-Scholes equation with constant coefficients $r$ , $q$  and $\sigma$ into $\bar{r}(t,T), \bar{q}(t,T)$ and $\overline{\sigma^2}(t,T)$. Using this fact and the proposition 1 at page 249 in [11], we soon have (2.7). A way of *direct proof* is as follows. As in [8], in (2.1) we use the changes of variable and unknown function $y = xe^{\bar{r}(t,T)-\bar{q}(t,T)}, U(y,t) = V(x,t)e^{\bar{r}(t,T)}$. Then (2.1) is changed to

$$\begin{cases} \dfrac{\partial U}{\partial t} + \dfrac{1}{2}\sigma^2(t)y^2\dfrac{\partial^2 U}{\partial y^2} = 0, & 0 < t < T, \quad y > 0, \\ U(y,\ T) = f(y), & y > 0. \end{cases}$$

If we change time variable into $\tau = \int_0^t \sigma^2(s)ds,\ \hat{T} = \int_0^T \sigma^2(s)ds$, then we have

$$\begin{cases} \dfrac{\partial U}{\partial \tau} + \dfrac{1}{2}y^2\dfrac{\partial^2 U}{\partial y^2} = 0, & 0 \le \tau < \hat{T}, \quad y > 0, \\ U(y,\hat{T}) = f(y), & y > 0. \end{cases}$$

This is the Black-Scholes equation with constant coefficients 0, 0 and 1 and thus we apply the proposition 1 at page 249 in [11] to get the representation of $U(y,\tau)$. Returning to original variables and unknown function, we get (2.7). (QED)

**Theorem 1**. (The Pricing Formulae of Higher Order Binary Options with Time Dependent Coefficients) *The prices of higher order bond and asset binaries with* risk free short rate $r(t)$, dividend rate $q(t)$ *and volatility* $\sigma(t)$ *are as follows.*

$$A_K^s(x,\ t\ ;\ T\ ;r(\cdot),q(\cdot),\sigma(\cdot)) = xe^{-\bar{q}(t,T)}N(sd),$$
$$B_K^s(x,\ t\ ;\ T\ ;r(\cdot),q(\cdot),\sigma(\cdot)) = e^{-\bar{r}(t,T)}N(sd\,'),\ \ s = + \text{ or } - \quad\quad (2.9)$$

$$N(x) = (\sqrt{2\pi})^{-1}\int_{-\infty}^x \exp[-y^2/2]dy,$$

$$d = \left(\sqrt{\overline{\sigma^2}(t,T)}\right)^{-1}\left[\ln(x/K)+\bar{r}(t,T)-\bar{q}(t,T)+\overline{\sigma^2}(t,T)/2\right],\ d\,' = d - \sqrt{\overline{\sigma^2}(t,T)},$$

$$A_{K_1K_2}^{s_1 s_2}(x,t\ ;T_1,T_2) = xe^{-\bar{q}(t,T_2)}N_2(s_1 d_1,\ s_2 d_2;\ s_1 s_2\rho),$$
$$B_{K_1K_2}^{s_1 s_2}(x,t\ ;T_1,T_2) = e^{-\bar{r}(t,T_2)}N_2(s_1 d_1',\ s_2 d_2';s_1 s_2\rho),\ \ \ s_1,s_2 = + \text{ or } -, \quad (2.10)$$

$$N_2(a,b\,;\rho) = \int_{-\infty}^a\int_{-\infty}^b \left(2\pi\sqrt{1-\rho^2}\right)^{-1}e^{-\frac{y^2-2\rho yz+z^2}{2(1-\rho^2)}}\,dydz,\quad \rho = \sqrt{\overline{\sigma^2}(t,T_1)\Big/\overline{\sigma^2}(t,T_2)},$$

$$A_{\bar{s_1}\cdots\bar{s_m}}^{s_1\cdots s_m}(x,t\ ;T_1,\cdots,T_m) = xe^{-\bar{q}(t,T_m)}N_m(s_1 d_1,\cdots,s_m d_m\ ;A_{s_1\cdots s_m}),s_i = + \text{ or } -, m \ge 3,$$
$$B_{\bar{s_1}\cdots\bar{s_m}}^{s_1\cdots s_m}(x,t\ ;T_1,\cdots,T_m) = e^{-\bar{r}(t,T_m)}N_m(s_1 d_1',\cdots,s_m d_m'\ ;A_{s_1\cdots s_m}),\ i = 1,\cdots,m\ , \quad (2.11)$$

$$N_m(a_1,\cdots,a_m\,;A) = \int_{-\infty}^{a_1}\cdots\int_{-\infty}^{a_m}\frac{1}{(\sqrt{2\pi})^m}\sqrt{\det A}\,\exp(-\frac{1}{2}\,y^{\perp}Ay)dy,$$

$$d_i = \left(\sqrt{\overline{\sigma^2}(t,T_i)}\right)^{-1}\left[\ln(x/K_i)+\overline{r}(t,T_i)-\overline{q}(t,T_i)+\overline{\sigma^2}(t,T_i)/2\right],$$

$$d_i{}' = d_i - \sqrt{\overline{\sigma^2}(t,T_i)}\,,\quad i=1,\cdots,m,$$

$$A_{s_1\cdots s_m} = (s_i s_j a_{ij})_{i,j=1}^m.$$

*Here* $y^{\perp}=(y_1,\cdots,y_m)$ *and the matrix* $(a_{ij})_{i,j=1}^m$ *is given as follows*:

$$a_{11}=\overline{\sigma^2}(t,T_2)\Big/\overline{\sigma^2}(T_1,T_2),\quad a_{mm}=\overline{\sigma^2}(t,T_m)\Big/\overline{\sigma^2}(T_{m-1},T_m),$$

$$a_{ii}=\overline{\sigma^2}(t,T_i)\Big/\overline{\sigma^2}(T_{i-1},T_i)+\overline{\sigma^2}(t,T_i)\Big/\overline{\sigma^2}(T_i,T_{i+1}),\quad 2\le i\le m-1,$$

$$a_{i,i+1}=a_{i+1,i}=-\sqrt{\overline{\sigma^2}(t,T_i)\cdot\overline{\sigma^2}(t,T_{i+1})}\Big/\overline{\sigma^2}(T_i,T_{i+1}),\ 1\le i\le m-1,$$

$$a_{ij}=0\ \text{for another}\ i,j=1,\cdots,m.$$

**Proof**. $A_K^s(x,\,t\,;\,T)$ is just the solution to the problems (2.1) and (2.6) when $f(x)=x\cdot 1\{sx>sK\}$. If we substitute $f(x)=x\cdot 1\{sx>sK\}$ into the second formula of (2.7), we soon get the formula for $A_K^s(x,\,t\,;\,T)$ of (2.9). Similarly, if we substitute $f(x)=1\{sx>sK\}$ into the first formula of (2.7), we soon get the formula for $B_K^s(x,\,t\,;\,T)$ of (2.9).

$A_{K_1K_2}^{s_1\,s_2}(x,t\,;T_1,T_2)$ is just the solution to the problems (2.1) and (2.6) when $T=T_1$ and

$f(x)=A_{K_2}^{s_2}(x,T_1;T_2)\cdot 1\{s_1x>s_1K_1\}$. If we substitute $f(x)=A_{K_2}^{s_2}(x,T_1;T_2)\cdot 1\{s_1x>s_1K_1\}$ and the

formula for $A_{K_2}^{s_2}(x,T_1;T_2)$ of (2.9) into the second formula of (2.7) and represent the integral with the cumulative distribution function of bivariate normal distribution, we get the formula for $A_{K_1K_2}^{s_1\,s_2}(x,t\,;T_1,T_2)$ of (2.10). Similarly, if we substitute $f(x)=B_{K_2}^{s_2}(x,T_1;T_2)\cdot 1\{s_1x>s_1K_1\}$ and the formula for $B_{K_2}^{s_2}(x,T_1;T_2)$ of (2.9) into the first formula of (2.7) and represent the integral with the cumulative distribution function of bivariate normal distribution, we get the formula for $B_{K_1K_2}^{s_1\,s_2}(x,t\,;T_1,T_2)$ of (2.10).

In the case of $m>2$, we use induction to prove (2.11). Assume that (2.11) holds for $m=$

$n-1$. Then from (2.4) $A_{\xi_1\xi_2\cdots\xi_n}^{s_1s_2\cdots s_n}(x,t\,;T_1,T_2,\cdots,T_n)$ satisfies (2.1) when $T=T_1$ and

$$V(x,T_1)=A_{\xi_2\cdots\xi_n}^{s_2\cdots s_n}(x,T_1;T_2,\cdots,T_n)\cdot 1(s_1x>s_1\xi_1)\,.$$

Then from the second formula of (2.7), $A_{\xi_1 \xi_2 \cdots \xi_n}^{s_1 s_2 \cdots s_n}(x,t;T_1,T_2,\cdots,T_n)$ is provided as follows:

$$A_{\xi_1 \xi_2 \cdots \xi_n}^{s_1 s_2 \cdots s_n}(x,t;T_1,T_2,\cdots,T_n) =$$

$$= xe^{-\bar{q}(t,T_1)}\int_0^{\infty}\frac{1}{\sqrt{2\pi\bar{\sigma}^2(t,T_1)}}\frac{1}{z^2}e^{-\frac{\left(\ln\frac{x}{z}+(\bar{r}(t,T_1)-\bar{q}(t,T_1)+\frac{1}{2}\bar{\sigma}^2(t,T_1)\right)^2}{2\bar{\sigma}^2(t,T_1)}}A_{\xi_2 \cdots \xi_n}^{s_2 \cdots s_n}(z,T_1;T_2,\cdots,T_n)\cdot 1(s_1 z > s_1\xi_1)dz.$$

Here $A_{\xi_2 \cdots \xi_n}^{s_2 \cdots s_n}(z,T_1;T_2,\cdots,T_n)$ is the price of the underlying $(n-1)$-th order asset binary option and thus by induction-assumption and (2.11) we have

$$A_{\xi_2 \cdots \xi_n}^{s_2 \cdots s_n}(z,T_1;T_2,\cdots,T_n) = ze^{-\bar{q}(T_1,T_n)}N_{n-1}(s_2 d_2,\cdots,s_n d_n;A_{s_2 \cdots s_n}).$$

If we substitute this equality into the above integral representation and represent the integral with the cumulative distribution function of $n$-variate normal distribution, we can get the first formula of (2.11) for $m = n$. The result for bond binaries (the second formula of (2.11)) is similarly proved. (QED.)

**Remark 1**. The theorem 1 is a generalization of the corresponding results of [5, 11]. In theorem 1, $N_2(a,b;\rho)$ is the *cumulative distribution function* of *bivariate normal distribution* with a *mean* vector $[0, 0]$ and a *covariance* matrix $[1, \rho; \rho, 1]$ (symbols in the software "Matlab"), and $N_m(a_1,\cdots,a_m;A)$ is the *cumulative distribution function* of *m-variate normal distribution* with zero *mean* vector and a *covariance* matrix $A^{-1} = (r_{ij})_{i,j=1}^m$ where $r_{ij} = (\sqrt{\bar{\sigma}^2(t,T_i)\Big/\bar{\sigma}^2(t,T_j)}$, $r_{ji} = r_{ij}, i \le j$. Such special functions can easily be calculated by standard functions supplied in standard software for mathematical calculation (for example, Matlab).

Next, we consider a *relation* between prices of higher order binaries with *constant difference between risk free rates and dividend rates*. From the formulae (2.9), (2.10) and (2.11), when $b$ is a constant, we have:

$$A_{K_1 \cdots K_m}^{s_1 \cdots s_m}(x,t;T_1,\cdots,T_m;r_1(\cdot),r_1(\cdot)+b,\sigma(\cdot)) = e^{-\overline{(r_1-r_2)}(t,T_m)}A_{K_1 \cdots K_m}^{s_1 \cdots s_m}(x,t;\cdots,T_m;r_2(\cdot),r_2(\cdot)+b,\sigma(\cdot)),$$

$$B_{K_1 \cdots K_m}^{s_1 \cdots s_m}(x,t;T_1,\cdots,T_m;r_1(\cdot),r_1(\cdot)+b,\sigma(\cdot)) = e^{-\overline{(r_1-r_2)}(t,T_m)}B_{K_1 \cdots K_m}^{s_1 \cdots s_m}(x,t;\cdots,T_m;r_2(\cdot),r_2(\cdot)+b,\sigma(\cdot)).$$

$$(2.12)$$

Next, as in [14], we can prove the following lemma. The proof is easy and omitted.

**Lemma 2.** (Integral of binary or nothing) *Assume that* $g(\tau)$ *is a continuous function of* $\tau \in [T_{i-1}, T]$. *Let*

$$V(x, T_0) = 1(s_0 x > s_0 \xi_0^i) \int_{T_{i-1}}^{T} g(\tau) \cdot F_{\xi_1 \cdots \xi_{i-1} \xi_i}^{s_1 \cdots s_{i-1} s_i} (x, \ T_0 \ ; T_1, \cdots, T_{i-1}, \tau) d\tau .$$  (2.13)

*Then the solution of (2.1) and (2.13) is given as follows:*

$$V(x, \ t) = \int_{T_{i-1}}^{T} g(\tau) \cdot F_{\xi_0 \xi_1 \cdots \xi_{i-1} \xi_i}^{s_0 s_1 \cdots s_{i-1} s_i} (x, \ t \ ; T_0, T_1, \cdots, T_{i-1}, \tau) d\tau .$$  (2.14)

*Here* $F = A \ or \ F = B$.

# 3. The Problem of Defaultable Bonds with Discrete Default Information, The Pricing Formulae and Credit Spread Analysis

## 3.1 The Problem

**Assumptions**: 1) Short rate follows the law

$$dr_t = a_r(r, \ t)dt + s_r(t)dW_1(t), \qquad a_r(r, \ t) = a_1(t) - a_2(t)r$$  (3.1)

under the risk neutral martingale measure and a standard Wiener process $W_1$.

2) $0 = t_0 < t_1 < \cdots < t_{N-1} < t_N = T$, $t_1, \cdots, t_N$ are announcing dates and $T$ is the maturity of our corporate bond with face value 1 (unit of currency). For every $i = 0, \cdots, N-1$ the unexpected default probability in $[t, \ t+dt] \bigcap [t_i, \ t_{i+1}]$ is $\lambda_i dt$. Here the *default intensity* $\lambda_i$ is a constant.

3) The firm value $V(t)$ follows a geometric Brown motion $dV(t) = (r_t - b)V(t)dt + s_V(t)V(t)dW_2(t)$ under the risk neutral martingale measure and a standard Wiener process $W_1$ and $E(dW_1, \ dW_2) = \rho dt$. The firm continuously pays out dividend in rate $b$ (constant) for a unit of firm value.

4) The expected default barrier is only given at time $t_i$ and the expected default event occurs when

$$V(t_i) \le K_i Z(r, t_i; T) \quad (i = 1, \cdots, N).$$

Here $K_i$ is a constant that *reflects the quantity of debt* and $Z(r, t; T)$ is default free zero coupon bond price.

5) The expected default recovery $R_{ed}$ is given by $R_e \cdot Z(r, t; T)$, the unexpected default recovery $R_{ud}$ as $R_u \cdot Z(r, t; T)$ and the *recovery rates* $0 \le R_e, R_u \le 1$ are constants. (**Exogenous** recovery.)

5)' The expected default recovery is given by $R_{ed} = R_e \cdot \alpha \cdot V$, the unexpected default recovery by $R_{ud} = \min\{Z(r, \ t), \ R_u \cdot \alpha \cdot V\}$ and the *recovery rates* $0 \le R_e, R_u \le 1$ are constants and $0 < \alpha < 1$ is a constant that *reflects the quantity of debt* (**Endogenous** recovery). Here the *reason* why the expected default recovery and unexpected recovery are *given* in *different* forms is to avoid the possibility of *paying more than the current price of risk free bond* as a default recovery when the unexpected default event occurs.

6) In the subinterval $(t_i, t_{i+1})$, the price of our corporate bond is given by a sufficiently smooth function $C_i(V, r, t)$ ($i = 0, \cdots, N-1$).

**Problem**: *Find the representation of the price function $C_i(V, r, t)$ ($i = 0, \cdots, N-1$) under the above assumptions.*

### 3.2 The Pricing Model

Under the assumption 1), the price $Z(r, t; T)$ of default free bond is the solution to the following problem

$$\begin{cases} \dfrac{\partial Z}{\partial t} + \dfrac{1}{2} s_r^2(t) \dfrac{\partial^2 Z}{\partial r^2} + a_r(r,t) \dfrac{\partial Z}{\partial r} - rZ = 0, \\ Z(r,T) = 1. \end{cases} \tag{3.2}$$

The solution is given by

$$Z(r, t; T) = e^{A(t,T) - B(t,T)r}. \tag{3.3}$$

Here $A(t, T)$ and $B(t, T)$ are differently given dependent on the specific model of short rate [16]. For example, if the short rate follows the *Vasicek* model, that is, if the coefficients $a_1(t)$, $a_2(t)$, $s_r(t)$ in (1) are all constants (that is, $a_1(t) \equiv a_1, a_2(t) \equiv a_2, s_r(t) \equiv s_r$), then $B(t, T)$ and $A(t, T)$ are respectively given as follows:

$$B(t,T) = \frac{1 - e^{-a_2(T-t)}}{a_2}, \quad A(t,T) = -\int_t^T \left[ a_2 B(u,T) - \frac{1}{2} s_r^2 B^2(u, T) \right] du. \tag{3.4}$$

See [16] for $B(t, T)$ and $A(t, T)$ in *Ho-Lee* model, *Hull-White* model and *CIR* model.

According to [16], under the above assumptions the price of defaultable bond with a constant default intensity $\lambda$ and unexpected default recovery $R_{ud}$ satisfies the following PDE:

$$\frac{\partial C}{\partial t} + \frac{1}{2} \left[ s_V^2(t)V^2 \frac{\partial^2 C}{\partial V^2} + 2\rho s_V(t)s_r(t)V \frac{\partial^2 C}{\partial V \partial r} + s_r^2(t) \frac{\partial^2 C}{\partial r^2} \right] +$$
$$+ (r-b)V \frac{\partial C}{\partial V} + a_r(r,t) \frac{\partial C}{\partial r} - (r+\lambda)C + \lambda R_{ud} = 0.$$

Therefore if we let $C_N(V, r, t) \equiv 1$ and , then the **price model** of our bond is given as follows:

$$\begin{cases} \dfrac{\partial C_i}{\partial t} + \dfrac{1}{2} \left[ s_V^2(t)V^2 \dfrac{\partial^2 C_i}{\partial V^2} + 2\rho s_V(t)s_r(t)V \dfrac{\partial^2 C_i}{\partial V \partial r} + s_r^2(t) \dfrac{\partial^2 C_i}{\partial r^2} \right] + (r-b)V \dfrac{\partial C_i}{\partial V} \\ \quad + a_r(r,t) \dfrac{\partial C_i}{\partial r} - (r+\lambda_i)C_i + \lambda_i R_{ud} = 0, \qquad t_i \le t < t_{i+1}, \\ C_i(t_{i+1}) = C_{i+1}(t_{i+1}) \cdot 1\{V > K_{i+1}Z\} + R_{ed} \cdot 1\{V \le K_{i+1}Z\}, \qquad i = 0, \cdots, N-1. \end{cases} \tag{3.5}$$

Here the default recoveries $R_{ed}$, $R_{ud}$ are differently given whether we choose to take the assumption 5) or 5)'. Under the assumption 5) the model is as follows:

$$\begin{cases} \dfrac{\partial C_i}{\partial t} + \dfrac{1}{2}\left[ s_V^2(t)V^2\dfrac{\partial^2 C_i}{\partial V^2} + 2\rho s_V(t)s_r(t)V\dfrac{\partial^2 C_i}{\partial V\partial r} + s_r^2(t)\dfrac{\partial^2 C_i}{\partial r^2}\right] + (r-b)V\dfrac{\partial C_i}{\partial V} \\ \qquad + a_r(r,t)\dfrac{\partial C_i}{\partial r} - (r+\lambda_i)C_i + \lambda_i R_u \cdot Z(r,t\,;T) = 0, \qquad t_i \le t < t_{i+1}, \\ C_i(t_{i+1}) = C_{i+1}(t_{i+1}) \cdot 1\{V > K_{i+1}Z\} + R_e \cdot Z(r,t\,;T)\cdot 1\{V \le K_{i+1}Z\}, i = 0,\cdots,N-1. \end{cases} \quad (3.6)$$

Under the assumption 5)' the model is as follows:

$$\begin{cases} \dfrac{\partial C_i}{\partial t} + \dfrac{1}{2}\left[ s_V^2(t)V^2\dfrac{\partial^2 C_i}{\partial V^2} + 2\rho s_V(t)s_r(t)V\dfrac{\partial^2 C_i}{\partial V\partial r} + s_r^2(t)\dfrac{\partial^2 C_i}{\partial r^2}\right] + (r-b)V\dfrac{\partial C_i}{\partial V} \\ \qquad + a_r(r,t)\dfrac{\partial C_i}{\partial r} - (r+\lambda_i)C_i + \lambda_i \min\{Z(r,t), R_u \cdot \alpha \cdot V\} = 0, \ t_i \le t < t_{i+1}, \\ C_i(t_{i+1}) = C_{i+1}(t_{i+1}) \cdot 1\{V > K_{i+1}Z\} + R_e \cdot \alpha \cdot V \cdot 1\{V \le K_{i+1}Z\}, \ i = 0,\cdots,N-1. \end{cases} \quad (3.7)$$

### 3.3   The Pricing Formulae

**Theorem 2.** (Exogenous recovery) *Under the assumptions 1), 2), 3), 4), 5) and 6), the price of our bond is represented as follows*:

$$C_i(V,r,t) = W_i(V/Z,t)\cdot Z + \left[1 - W_i(V/Z,t)\right]\cdot R_u \cdot Z, \ t_i \le \forall t < t_{i+1}, \ i = 0,\cdots,N-1. \quad (3.8)$$

*Here*

$$W_i(x,t) = e^{-\lambda_i(t_{i+1}-t)}\left\{ e^{-\sum\limits_{k=i+1}^{N-1}\lambda_k(t_{k+1}-t_k)} B_{K_{i+1}\cdots K_N}^{+\cdots+}(x,t\,;t_{i+1},\cdots,t_N\,;0\,,b\,,S_X(\cdot)) + \right.$$

$$\left. + \dfrac{R_e - R_u}{1 - R_u}\sum\limits_{m=i}^{N-1} e^{-\sum\limits_{k=i+1}^{m}\lambda_k(t_{k+1}-t_k)} B_{K_{i+1}\cdots K_m K_{m+1}}^{+\cdots+-}(x,t\,;t_{i+1},\cdots,t_m,t_{m+1}\,;0,b,S_X(\cdot))\right\}, \quad (3.9)$$

$$t_{N-2} \le t < t_{N-1}, \ x > 0, \ i = 0,\cdots,N-1,$$

$$S_X^2(t) = s_V^2(t) + 2\rho \cdot s_V(t)\cdot s_r(t)\cdot B(t,T) + [s_r(t)\cdot B(t,T)]^2 \ge 0. \quad (3.10)$$

*and* $B(t,T)$ *is given in (3.4);* $B_{K_1\cdots K_m}^{+\cdots+}(x,t\,;t_1,\cdots,t_m\,;0,b,s_X(\cdot))$ *is the price of m-th order bond binary with* 0-*risk free rate, b-dividend rate and* $S_X(t)$-*volatility.* (See the theorem 1.)

**Remark 2.** The theorem 2 is a generalization of the theorem 2 of [14] into the case of stochastic risk free rate. That is, if we let $r$ is a constant and $R_e = R_u$, we have the theorem 2 of [14]. The *financial meaning* of the pricing formulae (3.8) is very clear when $R = R_u = R_e$ and just the same as the one of the theorem 2 in [14]. $W_i(V/Z, t)$ is the *survival probability* after the time $t \in [t_i, t_{i+1})$, that is, the probability with which no default event occurs in the interval $[t, T]$ and $1 - W_i(V/Z, t)$ is the *ruin probability* after the time $t \in [t_i, t_{i+1})$, that is, the probability with which default event occurs in the interval $[t, T]$ when $t_i \le t < t_{i+1}$. The formulae (3.8) can be written as follows:

$$C_i(V, r, t) = R \cdot Z + (1-R)W_i(V/Z, t) \cdot Z, \ t_i \le \forall t < t_{i+1}, \ i = 0, \cdots, N-1. \quad (3.11)$$

The *financial meaning* of (3.11) is that the first term of (3.11) is the current price of the part to be given to bond holder *regardless of default occurs or not*, and the second term is the *allowance* dependent on the survival probability after time *t*.

**Theorem 3.** (Endogenous recovery)  *Under the assumptions 1), 2), 3), 4), 5)' and 6), the price of our bond is provided as follows*:

$$C_i(V, r, t) = Z(r, t) \cdot u_i(V/Z(r,t), t), \ t_i \le t < t_{i+1}, \ i = 0, \cdots, N-1. \quad (3.12)$$

*Here*

$$u_i(x, t) = e^{-\lambda_i(t_{i+1}-t)} \left\{ e^{-\sum_{k=i+1}^{N-1} \lambda_k(t_{k+1}-t_k)} B_{K_{i+1}\cdots K_N}^{+\cdots+}(x, t; t_{i+1}, \cdots, t_N; 0, b, S_X(\cdot)) + \right.$$

$$+ R_e \alpha \sum_{m=i}^{N-1} e^{-\sum_{k=i+1}^{m} \lambda_k(t_{k+1}-t_k)} A_{K_{i+1}\cdots K_m K_{m+1}}^{+\cdots+\ -}(x, t; t_{i+1}, \cdots, t_m, t_{m+1}; 0, b, S_X(\cdot))$$

$$+ \sum_{m=i+1}^{N-1} \lambda_m e^{-\sum_{k=i+1}^{m-1} \lambda_k(t_{k+1}-t_k)} \int_{t_m}^{t_{m+1}} e^{-\lambda_m(\tau-t_m)} \left[ B_{K_{i+1}\cdots K_m \frac{1}{R_u\alpha}}^{+\cdots+\ +}(x, t; t_{i+1}, \cdots, t_m, \tau; 0, b, S_X(\cdot)) + \right.$$

$$\left. + R_u \cdot \alpha \cdot A_{K_{i+1}\cdots K_m \frac{1}{R_u\alpha}}^{+\cdots+\ -}(x, t; t_{i+1}, \cdots, t_m, \tau; 0, b, S_X(\cdot)) \right] d\tau \left. \right\}$$

$$+ \lambda_i \int_{t}^{t_{i+1}} e^{-\lambda_i(\tau-t)} \left[ B_{\frac{1}{R_u\alpha}}^{+}(x, t; \tau; 0, b, S_X(\cdot)) + R_u \cdot \alpha \cdot A_{\frac{1}{R_u\alpha}}^{-}(x, t; \tau; 0, b, S_X(\cdot)) \right] d\tau. \quad (3.13)$$

$S_X^2(t)$ and $B(t,T)$ are given in (3.10) and (3.4); $B_{K_1\cdots K_m}^{+\cdots+}(x, t; t_1, \cdots, t_m; 0, b, s_X(\cdot))$ and $A_{K_1\cdots K_{m-1}K_m}^{+\cdots+\ -}(x, t; t_1, \cdots, t_{m-1}, t_m; 0, b, s_X(\cdot))$ are the prices of m-th order bond and asset binaries with 0-risk free rate, b-dividend rate and $S_X(t)$-volatility. (See the theorem 1.)

**Remark 3.** The theorem 3 is a generalization of the theorem 1 of [14] into the case of stochastic risk free rate. That is, if we let $r$ is a constant, $\alpha = 1/n$ and $\boldsymbol{R}_e = \boldsymbol{R}_u$, then we have the theorem 1, i) of [14]. The *financial meaning* of $u_i(x, t)$ is that it's the *relative price* of our bond in a subinterval *with respect to the risk free zero coupon bond*.

### 3.4    Credit Spread Analysis

In this subsection, we will illustrate the effect of several parameters including recovery rate, volatility of firm value, the relative price of the firm value and etc. on credit spreads. The *credit spread* is defined as the difference between the yields of defaultable bond $C$ and default-free bond $Z$ and is given by the following expression:

$$CS = -\frac{\ln C - \ln Z}{T - t}.$$

In the case of exogenous recovery (considered in theorem 2), the credit spread feature is similar with that of [14]. Here we consider the case of endogenous recovery (considered in theorem 3). In this case, the credit spread is differently given in every subinterval.

$$CS_i = -\frac{\ln(C_i(V, r, t)/Z(r, t))}{T - t} = -\frac{\ln u_i(V/Z(r, t), t)}{T - t}, \ t_i \le \forall t < t_{i+1}, \ i = 0, \cdots, N-1 \qquad (3.14)$$

Let $N = 2, \ t_1 = 3, \ t_2 = T = 6 \,(\text{annum})$

Basic data for calculation of $\boldsymbol{CS}$ is as follows: short rate model parameters: $a_1(t) \equiv 0.379*0.098$, $a_2(t) \equiv 0.379$, $s_r(t) \equiv 0.077$ (Vasicek model); firm value process parameters: dividend rate $b = 0.05$, volatility $s_V = 1.0$; $x = V/Z = 200$; correlation of short rate and firm value: $\rho = 0.5$; $\lambda_0 = 0.1$, $\lambda_1 = 0.3$ are respectively default intensity in the intervals $[0, t_1]$, $[t_1, t_2]$; $K_1 = K_2 = 100$ is default barrier at time $t_1, t_2$; recovery rate: $R_e = R_u = 0.5$; $\alpha = 1/150$.

We will analyze $(t : CS)$ plot changing one of $R, s_V, \ \rho, x = V/Z, \ \lambda$ and $K$ under keeping the remainder of data on as the above.

In what follows, the figure 1 shows that increase of recovery rate results in decrease of credit spread. The figure 2 shows that increase of volatility of firm value results in increase of credit spread. The reason is that when $s_V$ increases, the firm value fluctuates more seriously and there are more risks of default, which results in increase of credit spread. The figure 3 shows that increase of correlation between firm value and short rate results in increase of credit spread. The figure 4 shows that increase of firm value results in decrease of credit spread. The figures 5, 6 and 7 show that in the time interval close to the maturity increase of the default intensity results in increase of credit spread but in other time region the circumstance is not so simple. The figures 8, 9 and 10 show that the effect of default barrier on credit spread is different in the time intervals $[0, t_1]$ and $[t_1, T]$. The reason of such a complexity of the effect of default intensity and default barrier is in the formula (3.13).

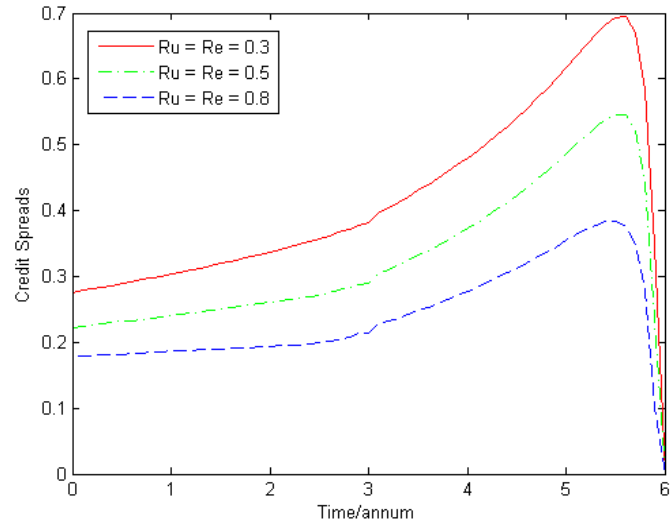

Figure 1. Plot ($t : CS$) when $R_e = R_u = 0.3, 0.5, 0.8$

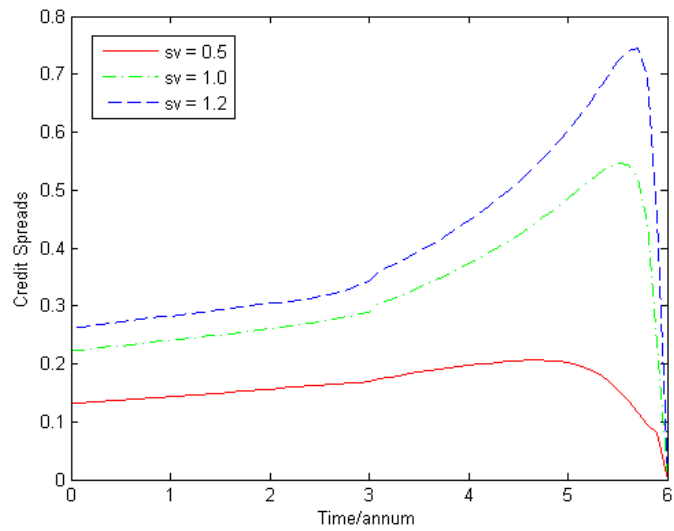

Figure 2. Plot ($t : CS$) when $s_V = 0.5, 1.0, 1.2$

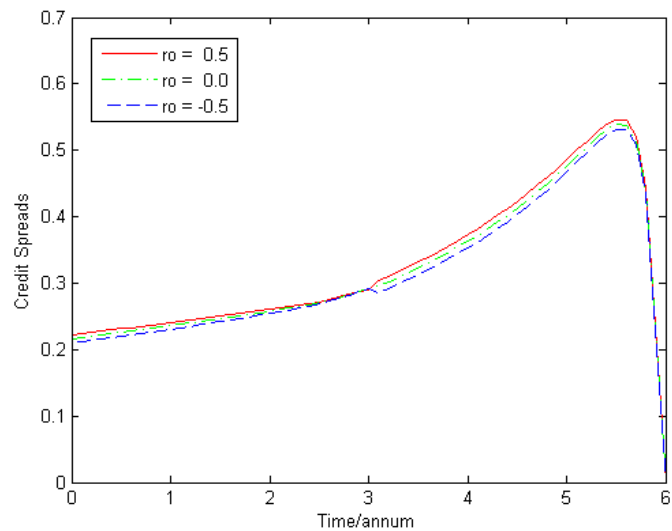

Figure 3. Plot ($t : CS$) when $\rho = 0.5, 0, -0.5$

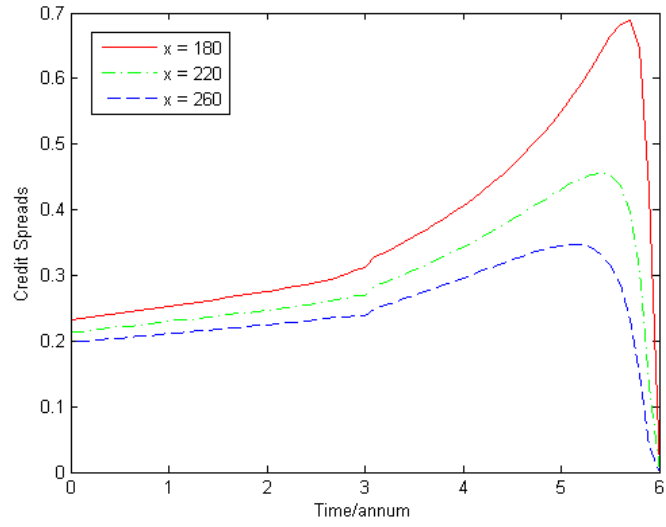

Figure 4. Plot ($t : CS$) when $x = V/Z = 180, 220, 260$

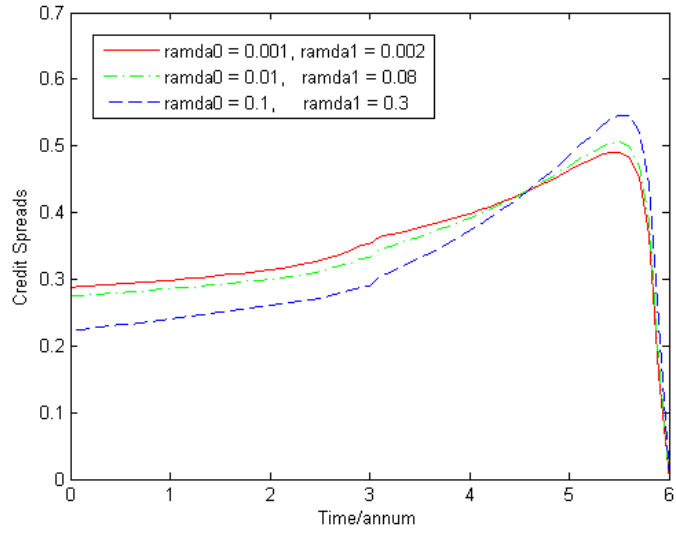

Figure 5. Plot ($t : CS$) when ($\lambda_0, \lambda_1$) = (0.001, 0.002), (0.01, 0.008), (0.1, 0.3)

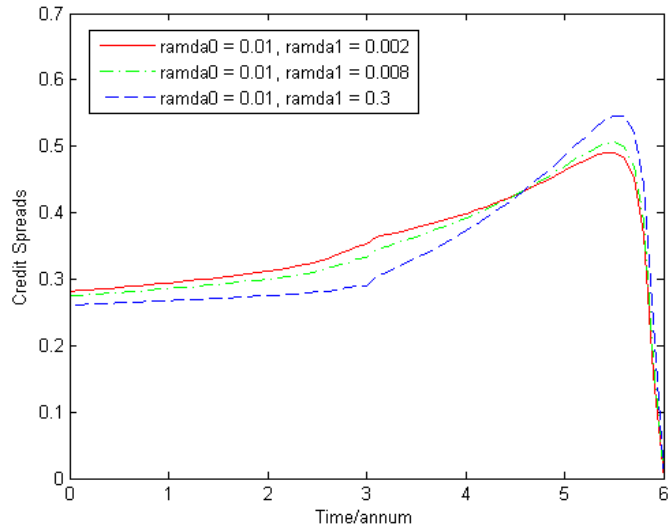

Figure 6. Plot ($t : CS$) when $\lambda_0 = 0.01$; $\lambda_1 = 0.002, 0.008, 0.3$

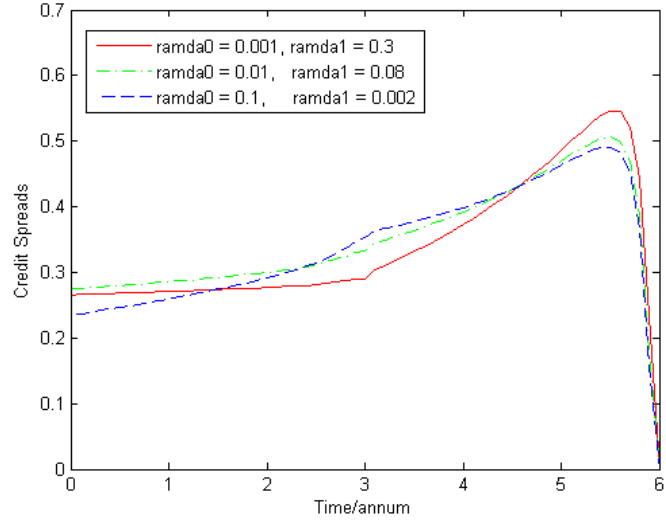

Figure 7. Plot ($t : CS$) when ($\lambda_0$, $\lambda_1$) = (0.001, 0.3), (0.01, 0.008), (0.1, 0.002)

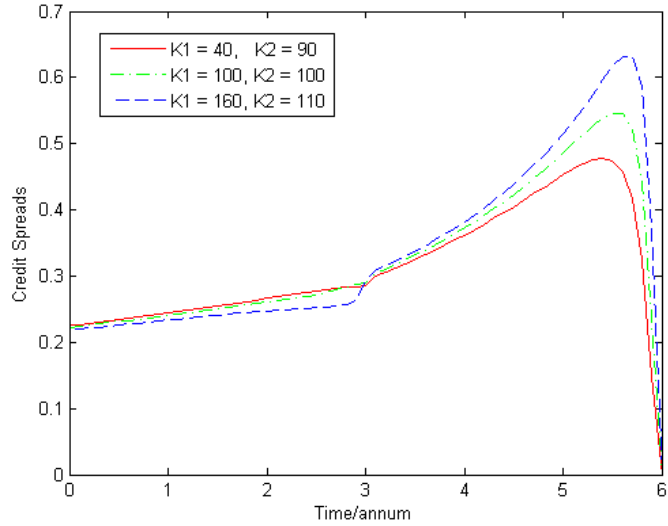

Figure 8. Plot ($t : CS$) when ($K_1$, $K_2$) = (40, 90), (100, 100), (160, 110)

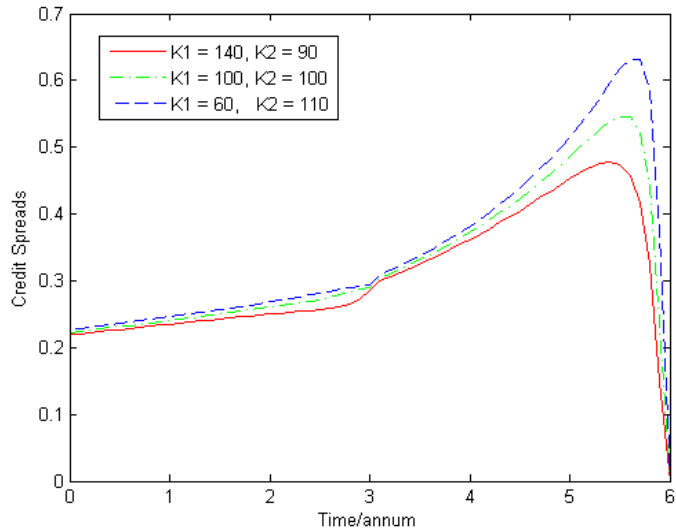

Figure 9. Plot ($t : CS$) when ($K_1$, $K_2$)= (140, 90), (100, 100), (60, 110)

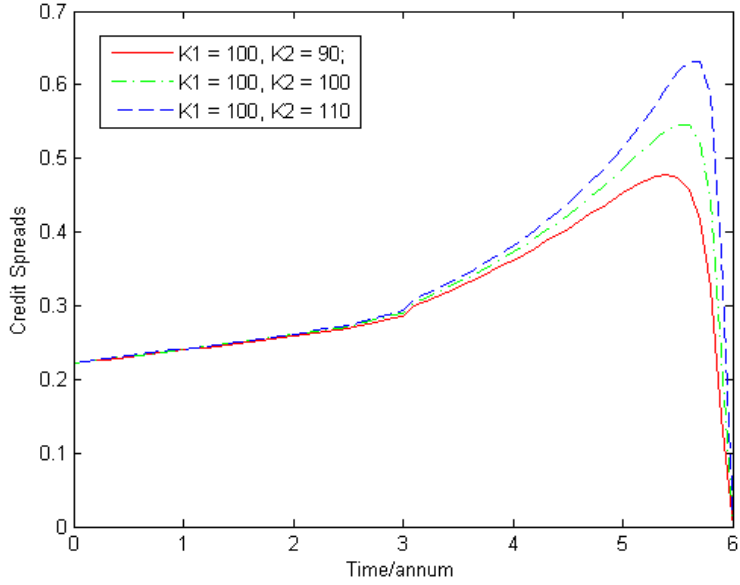

Figure 10. Plot ($t : CS$) when $K_1 = 100$; $K_2 = 90, 100, 110$

## 4.    The Proofs of The Pricing Formulae

**The Proof of Theorem 2**. Under the assumptions 1), 2), 3), 4), 5) and 6), the ***price model*** of our bond is given by (3.6). In (3.6), we use change of numeraire

$$x = V / Z(r,t), \quad u_i(x,t) = C_i(V, \ r, \ t)/Z(r,t), \ t_i \le t < t_{i+1}, \ i = 0,\cdots,N-1 . \qquad (4.1)$$

Here $\mathbf{Z}(\mathbf{r}, \mathbf{t})$ is the price of default free zero coupon bond given by (3.3). If we substitute (4.1) into the first equation of (3.6), note that

$$C_t = u_t Z - xu_x Z_t + uZ_t, \quad VC_V = xu_x Z, \quad C_r = Z_r(u - xu_x), \quad V^2 C_{VV} = x^2 u_{xx} Z,$$

$$VC_{Vr} = -x^2 u_{xx} Z_r, \qquad C_{rr} = Z_{rr}(u - xu_x) + x^2 u_{xx} Z_r^2 / Z, \quad Z_r / Z = -B(t,T)$$

and consider the equation (3.2) on $\mathbf{Z}(\mathbf{r}, \mathbf{t})$, then we have

$$u_t Z + \frac{1}{2} x^2 u_{xx} Z[s_V^2(t) + 2\rho s_V(t)s_r(t)B(t) + \left(s_r(t)B(t)\right)^2] - bxu_x Z - \lambda_i uZ + \lambda_i R_u Z(r,t) = 0 .$$

Divide the two hands by $\mathbf{Z}$ and let $S_X^2(t) = s_V^2(t) + 2\rho s_V(t) \cdot s_r(t) \cdot B(t) + \left(s_r(t)B(t)\right)^2$, then the problem (3.6) is changed to the following one dimensional problem:

$$\begin{cases} \dfrac{\partial u_i}{\partial t} + \dfrac{1}{2} S_X^2(t)x^2 \dfrac{\partial^2 u_i}{\partial x^2} - bx\dfrac{\partial u_i}{\partial x} - \lambda_i(u_i - R_u) = 0, & t_i < t < t_{i+1}, x > 0, \\ u_i(x,t_{i+1}) = u_{i+1}(x,t_{i+1}) \cdot 1\{x > K_{i+1}\} + R_e \cdot 1\{x \le K_{i+1}\}, & x > 0, \ i = 0,\cdots N-1, \end{cases} \qquad (4.2)$$

Here $u_N(x, \ t) \equiv 1$. We use the change of unknown function

$$u_i \ = \ (1 - R_u)W_i + R_u , \ (i = 0, \ 1, \ \cdots, \ N-1) \qquad (4.3)$$

to have

$$\begin{cases} \dfrac{\partial W_i}{\partial t} + \dfrac{1}{2} S_X^2(t) x^2 \dfrac{\partial^2 W_i}{\partial x^2} - bx \dfrac{\partial W_i}{\partial x} - \lambda_i W_i = 0, & t_i \le t < t_{i+1}, \ x > 0, \\ W_i(x,\ t_{i+1}) = W_{i+1}(x,\ t_{i+1}) \cdot 1\{x > K_{i+1}\} + \dfrac{R_e - R_u}{1 - R_u} \cdot 1\{x < K_{i+1}\}, & x > 0, \quad i = 0, \cdots N - 1. \end{cases} \quad (4.4)$$

Here $W_N(x, t) \equiv 1$. Then using this $W_i$, our bond price is provided by (3.8). The equation (4.4) is called the *equation* for the *survival probability* after the time $t \in [t_i,\ t_{i+1})$.

(4.4) is *a set* of Black-Scholes equations just like (4.22) in [14], but here the coefficient $S_X^2(t)$ is not a constant. So we use our theorem 1 instead of the theorems of [5, 11, and 12].

Now we solve the problem (4.4). When $i = N - 1$, (4.4) is

$$\begin{cases} \dfrac{\partial W_{N-1}}{\partial t} + \dfrac{1}{2} S_X^2(t) x^2 \dfrac{\partial^2 W_{N-1}}{\partial x^2} - bx \dfrac{\partial W_{N-1}}{\partial x} - \lambda_{N-1} W_{N-1} = 0, & t_{N-1} \le t < t_N, \ x > 0, \\ W_{N-1}(x,\ t_N) = 1\{x > K_N\} + \dfrac{R_e - R_u}{1 - R_u} \cdot 1\{x < K_N\}, & x > 0. \end{cases} \quad (4.5)$$

This is a pricing problem of binary options with coefficients $\lambda_{N-1}, \lambda_{N-1} + b, \ S_X(t)$ whose expiry payoff is a combination of bond and asset binaries. By the definition of bond binary, we have

$$W_{N-1}(x,t) = B_{K_N}^+(x,t;t_N;\lambda_{N-1},\lambda_{N-1} + b, S_X(\cdot)) + \dfrac{R_e - R_u}{1 - R_u} \cdot B_{K_N}^-(x,t;t_N;\lambda_{N-1},\lambda_{N-1} + b, S_X(\cdot)), \quad (4.6)$$
$$t_{N-1} \le t < t_N, \ x > 0.$$

Here $B_K^s(x,\ t;\ T;\ r(\cdot), q(\cdot), \sigma(\cdot))$ is given by the formula (2.9) of the theorem 1.

For our further purpose, using the relations (2.12) we rewrite (4.6) by the prices of bond and asset binaries with the coefficients $r = 0, \ q = b, \ \sigma(t) = S_X(t)$:

$$W_{N-1}(x,t) = e^{-\lambda_{N-1}(t_N - t)} B_{K_N}^+(x,t;t_N;0,b,S_X(\cdot)) + \dfrac{R_e - R_u}{1 - R_u} \cdot e^{-\lambda_{N-1}(t_N - t)} B_{K_N}^-(x,t;t_N;0,b,S_X(\cdot)), \quad (4.7)$$
$$t_{N-1} \le t < t_N, \ x > 0.$$

In order to solve (4.4) when $i = N - 2$, we need to rewrite (4.6) by the prices of bond and asset binaries with the coefficients $r = \lambda_{N-2}, \ q = \lambda_{N-2} + b, \ \sigma(t) = S_X(t)$ just as noted in the remark 3 in [14].

$$W_{N-1}(x,t) = e^{-(\lambda_{N-1} - \lambda_{N-2})(t_N - t)}[B_{K_N}^+(x,t;t_N;\lambda_{N-2}, \ \lambda_{N-2} + b, S_X(\cdot))$$
$$+ \dfrac{R_e - R_u}{1 - R_u} \cdot B_{K_N}^-(x,t;t_N;\lambda_{N-2}, \ \lambda_{N-2} + b, S_X(\cdot))], \ t_{N-1} \le t < t_N, \ x > 0. \quad (4.8)$$

When $i = N - 2$ using (4.8), (4.4) is written as follows:

$$\begin{cases} \dfrac{\partial W_{N-2}}{\partial t} + \dfrac{1}{2} S_X^2(t) x^2 \dfrac{\partial^2 W_{N-2}}{\partial x^2} - bx \dfrac{\partial W_{N-2}}{\partial x} - \lambda_{N-2} W_{N-2} = 0, \ t_{N-2} \le t < t_{N-1}, \ x > 0, \\ W_{N-2}(x,\ t_{N-1}) = e^{-(\lambda_{N-1} - \lambda_{N-2})(t_N - t_{N-1})}[B_{K_N}^+(x,t_{N-1};t_N;\lambda_{N-2}, \ \lambda_{N-2} + b, S_X(\cdot)) \cdot 1\{x > K_{N-1}\} + \\ + \dfrac{R_e - R_u}{1 - R_u} \cdot B_{K_N}^-(x,t;t_N;\lambda_{N-2}, \ \lambda_{N-2} + b, S_X(\cdot)) \cdot 1\{x > K_{N-1}\}] + \dfrac{R_e - R_u}{1 - R_u} \cdot 1\{x < K_{N-1}\}, \ x > 0. \end{cases} \quad (4.9)$$

This is a pricing problem of binary options with coefficients $\lambda_{N-2}$, $\lambda_{N-2} + b$, $S_X(t)$ whose expiry payoff is a combination of bond and asset binaries. By the definition of second order binary, we have

$$W_{N-2}(x,t) = e^{-(\lambda_{N-1} - \lambda_{N-2})(t_N - t_{N-1})}[B_{K_{N-1}K_N}^{+\ +}(x,t;t_{N-1},t_N;\lambda_{N-2},\ \lambda_{N-2}+b,S_X(\cdot))$$

$$+ \frac{R_e - R_u}{1 - R_u} \cdot B_{K_{N-1}K_N}^{+\ -}(x,t;t_{N-1},t_N;\lambda_{N-2},\ \lambda_{N-2}+b,S_X(\cdot))]+$$

$$+ \frac{R_e - R_u}{1 - R_u} B_{K_{N-1}}^{-}(x,t;t_{N-1};\lambda_{N-2},\ \lambda_{N-2}+b,S_X(\cdot)),\ t_{N-2} \le t < t_{N-1},\ x > 0.$$

Here $B_{K_1 K_2}^{s_1 s_2}(x,t;T_1,T_2;r(\cdot),q(\cdot),\sigma(\cdot))$ is given by the formula (2.10) of the theorem 1.

For our further purpose, using the relations (2.12) we rewrite $W_{N-2}(x,t)$ by the prices of bond and asset binaries with the coefficients $r = 0$, $q = b$, $\sigma(t) = S_X(t)$ :

$$W_{N-2}(x,t) = e^{-\lambda_{N-2}(t_{N-1}-t) - \lambda_{N-1}(t_N - t_{N-1})} B_{K_{N-1}K_N}^{+\ +}(x,t;t_{N-1},t_N;0,b,S_X(\cdot))$$

$$+ \frac{R_e - R_u}{1 - R_u} \cdot [e^{-\lambda_{N-2}(t_{N-1}-t) - \lambda_{N-1}(t_N - t_{N-1})} B_{K_{N-1}K_N}^{+\ -}(x,t;t_{N-1},t_N;0,b,S_X(\cdot))+ \qquad (4.10)$$

$$+ e^{-\lambda_{N-2}(t_{N-1}-t)} B_{K_{N-1}}^{-}(x,t;t_{N-1};0,b,S_X(\cdot))],\ t_{N-2} \le t < t_{N-1},\ x > 0.$$

By induction we have (3.9). Returning to original variables through (4.1) and (4.3), then we have the formula (3.8). (QED)

**The Proof of Theorem 3**. Under the assumptions 1), 2), 3), 4), 5)' and 6), the ***price model*** of our bond is given by (3.7). In (3.7), we use change of numeraire (4.1), then we have

$$\begin{cases} \dfrac{\partial u_i}{\partial t} + \dfrac{1}{2} S_X^2(t) x^2 \dfrac{\partial^2 u_i}{\partial x^2} - bx \dfrac{\partial u_i}{\partial x} - \lambda_i u_i + \lambda_i \min\{1,\ R_u \alpha \cdot x\} = 0,\ t_i < t < t_{i+1}, x > 0, \\ u_i(x,t_{i+1}) = u_{i+1}(x,t_{i+1}) \cdot 1\{x > K_{i+1}\} + R_e \alpha \cdot x \cdot 1\{x \le K_{i+1}\},\ x > 0,\ i = 0,\cdots N-1, \end{cases} \qquad (4.11)$$

The (4.11) is a similar problem with the problem (4.5) in [14]. The only difference is that the (4.11) is a set of terminal value problems for inhomogenous Black-Scholes equations with time dependent coefficients but the (4.5) in [14] is a set of terminal value problems for inhomogenous Black-Scholes equations with constant coefficients. If we follow the way of solving (4.5) in [14] using our theorem 1, lemma 2 and the relations (2.12), then we can get the formula (3.13). Then returning to the original variable ***V*** and the unknown function ***C*** using (4.1) we can soon obtain the formula (3.12). The detail is omitted. (QED)

# 5. Conclusions

1) We proved the pricing formula of *higher order binary option with time dependent coefficients* (theorem 1). This is a generalization of the corresponding results of [5, 11]. And we generalized the *integral formula* of higher order binary option on the last expiry date variable into the case with time dependent coefficients (lemma 2).

2) We obtained the pricing formulae of Two factor - model for defaultable bonds with discrete default intensity and discrete default barrier in both cases of exogenous and endogenous recoveries (theorem 2 and theorem 3) using the pricing formulae of higher order binary options with time dependent coefficients.

3) In further study the method can seemingly be applied to generalization of the study of [1] into the pricing defaultable coupon bond in combining model with the structural approach and the reduced form approach.

**Acknowledgement:** Authors thank anonymous arXiv moderators for strict note which helps to make this version better and more complete.